# All-optical diode based on plasmonic attenuation and nonlinear frequency conversion

Ming-Liang Ren, Xiao-Lan Zhong, Bao-Qin Chen and Zhi-Yuan Li*

*Laboratory of Optical Physics, Institute of Physics, Chinese Academy of Sciences, P.O. Box 603, Beijing 100190, China*

Abstract

We present design of an all-optical diode in a metal-dielectric structure where plasmonic attenuation and quasi-phase-matching (QPM) is harnessed to improve its performance greatly. Due to the anti-symmetric design of the nonlinear susceptibility, different incident direction will ignite different plasmonic nonlinear process, which either compensates plasmonic attenuation sufficiently or accelerates it seriously. As a result, unidirectional output of plasmonic signal is achieved. This designed all-optical diode shows advantages of low power consumption, short sample length, high isolation contrast, wide acceptance of structural and initial conditions, and tunable unidirectionality, and becomes of practical interest.

PACS numbers: 73.20.Mf, 42.65.Ky, 42.70.Mp

A diode that operates with photons instead of electrons has been an increasing interest due to its wide applications in integrated photonic circuits and information processing. It is associated with the breaking time-reversal symmetry of light, and is typically achieved in magneto-optical materials [1-3]. Optical nonlinearity is another useful routine to realize light signal processing [4,5] and optical isolator [6-9]. Recently, unidirectional light propagation has been experimentally observed in silicon photonic structures [10] and photonic crystal fiber [11]. For the sake of all-optical integration, it is highly desirable to design miniaturized optical diodes [10].

Plasmonics is a fascinating and rapidly expanding area of research as it offers a good resolution to achieve extremely small mode wavelengths and large enhancement of local electromagnetic field. It has enabled great potential for biological sensing and labeling, sub-wavelength revolution imaging, molecules emission and so on [12]. Giant enhancement of second order nonlinear interaction has been observed in plasmonic structures [13-15]. However, plasmonic attenuation due to intrinsic metal absorption will fundamentally limit the performance of plasmonic devices. In this work, we present design of an all-optical diode based on asymmetric second order nonlinear interaction in a plasmonic structure. Interestingly, plasmonic attenuation is no longer a disadvantageous obstacle, instead, it can be harnessed to shorten this device considerably. Moreover, the nonlinear interaction is facilitated greatly by the quasi-phase-matching (QPM) scheme and local field enhancement in this plasmonic structure [14], leading to a lower level of pump power.

The geometry of the designed diode is depicted in Fig. 1(a), where a silver film is deposited on the periodically poled lithium niobate (PPLN) substrate which starts from a positive domain and ends as a negative one. Surface plasmon polariton (SPP) at the frequencies of $\omega_1$ and $\omega_2$ (=$2\omega_1$) can be excited simultaneously in Ag-PPLN interface ($z$=0). Here we only consider nonlinear interaction between SPP modes for convenience, and assume SPP with $\omega_1$ as the fundamental wave (FW) and the one with $\omega_2$ as second harmonic wave (SHW). Since SHW gets attenuated more seriously than FW at the studied wavelength range [Fig. 1(b)], we choose FW as a pump or control light while SHW as a signal one, which will reveal the unidirectional transport.

This design is different from that of most light signal processing via quadratic nonlinearity where FW is usually controlled by seeded SHW and behaves as a signal light [4,5], and that of traditional PPLN diodes where a defect domain or layer is required [8,9].

In order to describe this device explicitly, we start from the coupled mode theory in waveguide, where the electromagnetic wave can be written as $E_l = \sum A_l(x) E_l(z) \exp(ik_l x)$, and $H_l = \sum A_l(x) H_l(z) \exp(ik_l x)$. Here $E_l(z)$ and $H_l(z)$ are mode profiles which have been normalized; $A_l(x)$ is the amplitude with $\omega_l$ ($l=1,2$) which can be described by the nonlinear coupled equations [13,15] as

$$\frac{dA_1}{dx} = -\frac{\alpha_1}{2} A_1 + i\frac{\omega\varepsilon_0}{4} \kappa_1 A_1^* A_2 e^{i\Delta\beta x}, \quad (1)$$

$$\frac{dA_2}{dx} = -\frac{\alpha_2}{2} A_2 + i\frac{\omega\varepsilon_0}{4} \kappa_2 A_1^2 e^{-i\Delta\beta x}, \quad (2)$$

Here $\Delta\beta = \beta_2 - 2\beta_1$. $\beta_l$ and $\alpha_l/2$ stand for the real and imaginary part of wave vector $k_l$, respectively, namely $k_l = \beta_l + i\alpha_l/2$ ($l=1,2$). For the nonlinear susceptibility $d_{33}(x,z)$ of PPLN is quite larger than that of silver and other elements of PPLN, such as $d_{31}(x,z)$, we only consider the nonlinear interaction due to $d_{33}(x,z)$ in PPLN substrate, and assume $d_{33}(x, z > 0) = 0$ and $d_{33}(x, z < 0) = d_{33}(x)$. Therefore the involved SPP modes should both polarize in $z$-axis. We define the nonlinear coupling coefficient as $\kappa = \int_{-\infty}^{0} E_{2z}^* (E_{1z})^2 dz$, which reveals the mode overlap in PPLN substrate. Therefore $\kappa_1$ in Eq. (1) and $\kappa_2$ in Eq. (2) can be expressed as $\kappa_1 = \int_{-\infty}^{+\infty} \chi^{(2)} : \vec{E}_2 \vec{E}_1^* \cdot \vec{E}_1^* dz = d_{33}(x)\kappa^*$ and $\kappa_2 = \int_{-\infty}^{+\infty} \chi^{(2)} : \vec{E}_1 \vec{E}_1 \cdot \vec{E}_2^* dz = d_{33}(x)\kappa$.

Since modulated periodically, the nonlinear coefficient can be written as Fourier series, $d_{33}(x) = \sum_m d_m e^{iG_m x}$, where $d_m = |d_{33}| e^{-i\delta_m} \sin(\pi m D)/(\pi m/2)$ and $G_m = m2\pi/\Lambda$ for PPLN with a duty cycle $D$ and periodicity $\Lambda$. $\delta_m = \pi - \pi m D$ for $d_{33} > 0$, while $\delta_m = \pi m D - \pi$ for $d_{33} < 0$. Assume $A_1 = |A_1| e^{i\phi_1}$, $A_2 = |A_2| e^{i\phi_2}$, and $\kappa = |\kappa| e^{i\phi_0}$. When $\Delta\beta = G_m$ or QPM is achieved, Eq. (1) and Eq. (2) can be revised as

$$\frac{d|A_1|}{dx} = \frac{1}{2}|A_1|[-\alpha_1 - \frac{1}{2}\omega\varepsilon_0|d_m||\kappa||A_2|\sin(\varphi)], \tag{3}$$

$$\frac{d|A_2|}{dx} = \frac{1}{2}|A_2|[-\alpha_2 + \frac{1}{2}\omega\varepsilon_0|d_m||\kappa|\frac{|A_1|^2}{|A_2|}\sin(\varphi)], \tag{4}$$

$$\frac{d\varphi}{dx} = \frac{1}{4}\omega\varepsilon_0|d_m||\kappa||A_2|(\frac{|A_1|^2}{|A_2|^2}-2)\cos(\varphi). \tag{5}$$

Here the effective nonlinear susceptibility model (ESM) is adopted [16,17], and the relative phase, $\varphi(x)=\phi_2(x)-2\phi_1(x)-\phi_0+\delta_m$, is determined by SHW, FW, the nonlinear coupling coefficient and domain sign. We call $\varphi(0)$ the initial relative phase, which can be utilized to estimate the energy transfer direction when both FW and SHW exist initially [18]. In particular, FW is directly converted into SHW in the case of $\varphi(0)=\pi/2$, while first gets amplified with $\varphi(0)=-\pi/2$. This is the phase-sensitive effect [4,5].

Assume the initial amplitude, $|A_{i0}|=|A_i(0)|$ ($i$=1,2). If $|A_{20}|<<|A_{10}|$, FW is rarely depleted and $|A_1(x)|\approx|A_{10}|\exp(-\alpha_1 x/2)$ holds true in the whole nonlinear interaction. Therefore, the SHW amplitude can be derived in the simple case of $\varphi_\pm(0)=\pm\pi/2$, $m$ =1 and $D$=0.5,

$$|A_2^{(\pm)}(L)| = |A_{20}|e^{-\gamma_2 L}\{1 \pm g[e^{(\gamma_2-\gamma_1)L}-1]\}, \tag{6}$$

where $g=\omega\varepsilon_0|d_1||\kappa||A_{10}|^2/[4|A_{20}|(\gamma_2-\gamma_1)]$, $\gamma_2=\alpha_2/2$, $\gamma_1=\alpha_1$, and $L$ is the sample length. For both FW and SHW are SPP modes, they have good mode overlap and large nonlinear coupling coefficient $|\kappa|$. Moreover, low QPM order ($m$ =1) is chosen for large effective nonlinear susceptibility $|d_m|$. These designs can facilitate nonlinear interaction and lower pump intensity. Notice that $\delta_1$ corresponds to the domain sign and appears to be different in the two directions due to the anti-symmetric structure. Therefore, different incident direction will lead to different transmission, even unidirectionality. Here the sign (+) or (-) stands for the forward or backward direction. Suppose $\eta_\pm = |A_2^{(\pm)}(L)|^2/|A_{20}|^2$ and define the isolation contrast $C$ to evaluate the performance of this diode,

$$C = \frac{\eta_+ - \eta_-}{\eta_+ + \eta_-} = \frac{2g[e^{(\gamma_2-\gamma_1)L}-1]}{1+\{g[e^{(\gamma_2-\gamma_1)L}-1]\}^2}. \quad (7)$$

Obviously, $C=\pm 1$ relates to the perfect unidirectionality. When $g[e^{(\gamma_2-\gamma_1)L}-1]=1$, we can obtain $\eta_+ = 4e^{-2\gamma_2 L}$, $\eta_- = 0$ and $C=1$. Note that $g$ is associated with structural parameters (e.g. $d_1$, $\kappa$, $\gamma_2$ and $\gamma_1$) and initial light conditions (e.g. $A_{10}$ and $A_{20}$). We can adjust structural parameters (e.g. $\gamma_2$ and $L$) to obtain $4e^{-2\gamma_2 L}=1$, and then control the initial light to achieve $g[e^{(\gamma_2-\gamma_1)L}-1]=1$. Thus we can achieve $\eta_+ = 1$, $\eta_- = 0$ and $C=1$. This is the unidirectionality based on plamonic attenuation and anti-symmetric nonlinear frequency conversion.

As an example to explicitly demonstrate how the above ideas work, we tune FW pump at 1.55 μm and SHW signal accordingly at 0.775 μm. The corresponding effective indices of these two SPP modes can be calculated, respectively $n_1$=2.17688+0.002655$i$ and $n_2$=2.38016+0.023157$i$. Each domain length of PPLN is set as the coherence length $l_c= \pi/\Delta\beta$ =1.906 μm, which can be achieved by the indirect electron-beam poling method [19] with the help of precise control. Assume the positive (negative) domain length is $l_p$ ($l_n$). Therefore $l_p=l_n=l_c$, $\Lambda=\Lambda_0=l_p+l_n=2l_c$ and $D=l_n/\Lambda$=0.5. For domain inversion only changes the nonlinear susceptibility ($d_{33}$) of lithium niobate (LN) instead of refractive index ($n$), the SPP modes in Ag-PPLN has little difference from that in Ag-LN. Furthermore, it is seen that $\gamma_2>\gamma_1$ and $g[e^{(\gamma_2-\gamma_1)L}-1]=1$ can come true easily.

Suppose two constant intensities $|A_{100}|^2$=1.44 MW/cm and $|A_{200}|^2$=816.3 W/cm. We first assume $|A_{10}|^2=|A_{100}|^2$ and $|A_{20}|^2=|A_{200}|^2$. When FW pump and SHW signal are incident upon the left side of the sample, they first "see" a positive domain and $\delta_1=\pi/2$. If the initial relative phase is tuned as $\pi/2$ or $\phi_2(0) - 2\phi_1(0) - \phi_0 = 0$, second harmonic generation (SHG) takes place and FW is converted into SHW gradually. As controlled by the initial FW and SHW, SHW converted from FW should be coherent with SHW input. Therefore, this process can be regarded as the nonlinear gain for SHW signal input. At the beginning, the nonlinear gain is large and can compensate plasmonic

attenuation sufficiently. Therefore, SHW signal seems to get amplified slightly [Fig. 2(a)]. Usually, this process could increase SHW signal continuously due to perfect quasi-phase-matching (QPM) scheme. However, FW pump also gets attenuated when traveling [Fig. 1(b)] and the nonlinear gain decreases accordingly. When it cannot compensate plasmonic attenuation, SHW signal starts to decline. Following Eq. (1) and Eq. (2), we also simulate this case numerically, and find the SHW signal evolution exhibits the same tendency with the analytical one. But in each domain, $\Delta\beta \neq 0$, and SHW signal seems to vibrate around the average line or analytical one [Fig. 2(a)] and exchanges the energy with FW pump periodically. The vibrating periodicity seems to get close to $\pi/\Delta\beta$.

If FW pump and SHW signal propagate backwardly with $\phi_2(0) - 2\phi_1(0) - \phi_0 = 0$, they first encounter a negative domain and $\delta_1 = -\pi/2$. In this situation, the initial relative phase becomes $-\pi/2$. Thus Optical parametric process first comes forth and SHW signal is converted into FW. Together with plasmonic attenuation, this process accelerates to consume SHW signal seriously which vanishes only at a distance of two-domain length, about 3.8 μm [Fig. 2(b)]. In contrast, SHW signal should travel more than ten domains in the case without nonlinearity [Fig. 1(b)] and get exhausted completely only due to plasmonic attenuation. When it decreases to zero, optical parametric process comes to end and then SHG is ignited, building SHW signal again. This is the cascaded second order nonlinear effect [19,20]. Similarly, the rebuilding SHW signal will approach the maximum and then decline gradually due to plasmonic attenuation of FW. The numerical result likewise reveals the vibration of SHW signal.

Considering both cases, we give a plot of the isolation contrast in Fig. 2(c). The perfect unidirectionality is seen to come true at an optimal distance of two-domain length where the forward SHW signal output nearly maintains its initial value [Fig. 2(a)] and the backward one is consumed completely [Fig. 2(b)]. Suppose the SHW signal output which remains (or vanishes) is ON (or OFF). Therefore the forward SHW output is ON while the backward one is OFF in this situation. In the following, we will still suppose $\phi_2(0) - 2\phi_1(0) - \phi_0 = 0$ and choose a pair of positive and negative domains as our device which is only about 3.8 μm long and four orders of magnitude smaller than the PPLN waveguide diode [8]. We believe that it will greatly facilitate its application in nanophotonic integration and attract other attention.

Usually, domain expansion is inevitable in the domain poling process, perhaps leading to a little shift of the duty circle ($D$) and periodicity ($\Lambda$). However, this device has a wide response to these factors [Fig. 3]. When $\Lambda=\Lambda_0$ (the white vertical dash line), the isolation contrast ($C$) is still high (>0.9) even if the duty circle varies from 0.4 to 0.6. This varying range is wide and can be realized easily in practice. Similarly, even if $\Lambda$ fluctuates from 0.9 $\Lambda_0$ to 1.1 $\Lambda_0$, the device still maintains $C$>0.9 when $D$=0.5 (the white horizontal dash line). It is also seen that a large periodicity is usually associated with a small duty circle to achieve high isolation contrast. For example, $\Lambda=1.1\Lambda_0$ (0.9$\Lambda_0$) and $D$=0.4 (0.6) corresponds to $C$=1.

We proceed to investigate this all-optical diode in different initial conditions. Figures 4(a)-(c) correspond to the initial FW pump with $|A_{20}|^2=|A_{200}|^2$. In the forward case, the SHG process occurs to offer the nonlinear gain and support the transport of SHW signal. At low intensity of the initial FW, the nonlinear gain is too small to compensate plasmonic attenuation sufficiently. Therefore, SHW signal gets consumed and its transmission drops below 1.0. With the increasing of the initial FW intensity, SHW signal output continues to blow up and then gets amplified. However, the optical parametric process first comes out in the backward case, considerably heightening SHW signal attenuation. Therefore it becomes unavailable at a certain (optimal) initial intensity ($|A_{100}|^2$) as utilized above. Further increasing the initial FW intensity will lead to the rebuilding of SHW signal, for the cascaded second order nonlinear effect occurs and SHG is aroused. In this direction, it is very interestingly seen that different initial pump intensity can result in different signal transmission [Fig. 4(b)]. Therefore one can also control the initial pump intensity to achieve optical switch based on cascaded second-order nonlinear process [4,5].

In addition to the initial FW pump, unidirectional light response is also studied against the initial SHW signal with $|A_{10}|^2=|A_{100}|^2$, and plotted in Figs. 4(e)-(f). For

$g \propto |A_{10}|^2 / |A_{20}|$ in Eq.(6), the forward SHW signal output decreases as the initial SHW signal input increases [Fig. 4(e)]. Furthermore, there also exists an optimal initial SHW intensity ($|A_{200}|^2$) at which the backward SHW is consumed completely [Fig. 4(f)] and the isolation contrast is high [Fig. 4(g)]. If SHW signal is initially zero, SHG would take place efficiently due to QPM scheme in both two directions and no unidirectionality comes true regardless of the anti-symmetric design. More interestingly, a logic gate of NOT SHW signal can be realized in the backward direction as long as FW pump exists. As is depicted in Table 1, SHW signal output is present due to the SHG process without SHW input. However it appears to be isolated in some conditions, such as $|A_{20}|^2=|A_{200}|^2$,

Furthermore, it is observed that this plasmonic device reveals a wide acceptance of initial intensity of both FW and SHW [Fig. 5], which is crucial for its practical applications. For $g \propto |A_{10}|^2 / |A_{20}|$ in Eq. (6), large initial SHW usually relates to large initial FW for high isolation contrast. This means one can always tune FW pump ($A_{10}$) for any given and unknown SHW signal ($A_{20}$) to achieve good unidirectionality. The vertical and horizontal white dash lines in Fig. 5 refer to Fig. 4(c) and Fig. 4(g), respectively.

As mentioned above, the initial relative phase plays a very important role in this diode. It is seen that SHW signal output oscillates with the initial relative phase periodically in Fig. 6. In each direction, ON and OFF states are switchable, leading to the phase-sensitive optical switch [4]. When $\varphi_\pm(0) = \pm\pi/2$, plasmonic attenuation is compensated for sufficiently in the forward direction, and SHW signal becomes ON. However, it appears to be OFF in the backward direction, for its all power is consumed seriously by the nonlinear interaction and plasmonic attenuation. In this situation, the isolation contrast gets close to 1.0. When $\varphi_\pm(0) = \mp \pi/2$, the unidirectionality still exists, but the ON direction is reversed and the isolation contrast becomes -1.0. This feature means that one can achieve each state (ON or OFF) in each direction by simply tuning the initial relative phase. To this point, the unidirectional behavior of this device is ascribed to the phase-sensitive effect in essence [4,5], and the different initial relative phase only originates from the anti-symmetric structure for all initial conditions of FW and SHW are fixed in two directions. In practice, SHW signal and Ag-PPLN structure are usually given but some of their parameters may be unknown, such as $|A_{20}|$, $\phi_2(0)$ and $\phi_0$. However, FW pump can be controlled flexibly and utilized to tune the initial relative phase (e.g. $\pm\pi/2$) and isolation contrast (e.g. $\pm 1$) conveniently. In this way, one can achieve not only tunable optical diode [9] but also optical switch [4,5].

In summary, we have exploited the all-optical diode based on plasmonic attenuation and nonlinear frequency conversion in the Ag-PPLN interface with an anti-symmetric PPLN configuration. Tunable unidirectionality can be achieved by means of external control conditions, such as FW pump here and acoustic wave [9]. Besides PPLN, this principle is also applicable to other semiconductors, such as GaAs which possesses large nonlinear susceptibility and is usually designed for the QPM scheme [22]. Moreover, one can also design a hybrid plasmonic structure [15] for SHW pump and control FW transmission. In general, this plasmonic device involves local field enhancement and QPM, and has many promising properties, including low power consumption, short sample length, high isolation contrast and so on. This will open up a new avenue towards practical application in information processing and nanophotonic integration.

This work was supported by the State Key Development Program for Basic Research of China at No. 2011CB922002 and the National Natural Science Foundation of China at Nos.10634080 and 60736041.

Corresponding authors: *renbright@aphy.iphy.ac.cn, lizy@aphy.iphy.ac.cn*

Figure captions:

Fig. 1. (Color online) (a) Schematic diagram of the Ag-PPLN structure and (b) calculated plasmonic attenuation of FW and SHW. $l_c$ is the coherence length and the Ag-PPLN interface is located at $z = 0$.

Fig. 2. (Color online) Analytical and numerical result for the SHW signal evolution as a function of propagation length. Panel (a) stands for the forward case, (b) represents the backward one, and (c) depicts the isolation contrast.

Fig. 3. (Color online) Isolation contrast calculated against the duty circle and periodicity according to Eq. (1) and Eq. (2). $L= \Lambda$, $\Lambda_0=2l_c$ and $\phi_2(0) - 2\phi_1(0) - \phi_0 = 0$.

Fig. 4. (Color online) SHW signal propagation characteristic studied in different pump conditions. Panels (a)-(c) correspond to FW pump intensity with $|A_{20}|^2=|A_{200}|^2$, while (e)-(g) are related to the initial intensity of SHW with $|A_{10}|^2=|A_{100}|^2$. $L=2l_c$, $D=0.5$, $|A_{100}|^2 =1.44$ MW/cm, $|A_{200}|^2=816.3$ W/cm and $\phi_2(0) - 2\phi_1(0) - \phi_0 = 0$.

Fig. 5. (Color online) Isolation contrast plotted against initial FW pump and SHW signal according to Eq. (1) and Eq. (2).

Fig. 6. (Color online) Unidirectional light response versus the initial relative phase according to Eq. (1) and Eq. (2). Here $|A_{10}|^2=|A_{100}|^2$, $|A_{20}|^2=|A_{200}|^2$, $L=2l_c$ and $D=0.5$.

Table 1 A logic gate of NOT SHW in the Ag-PPLN structure. "1" stands for being present while "0" for being absent

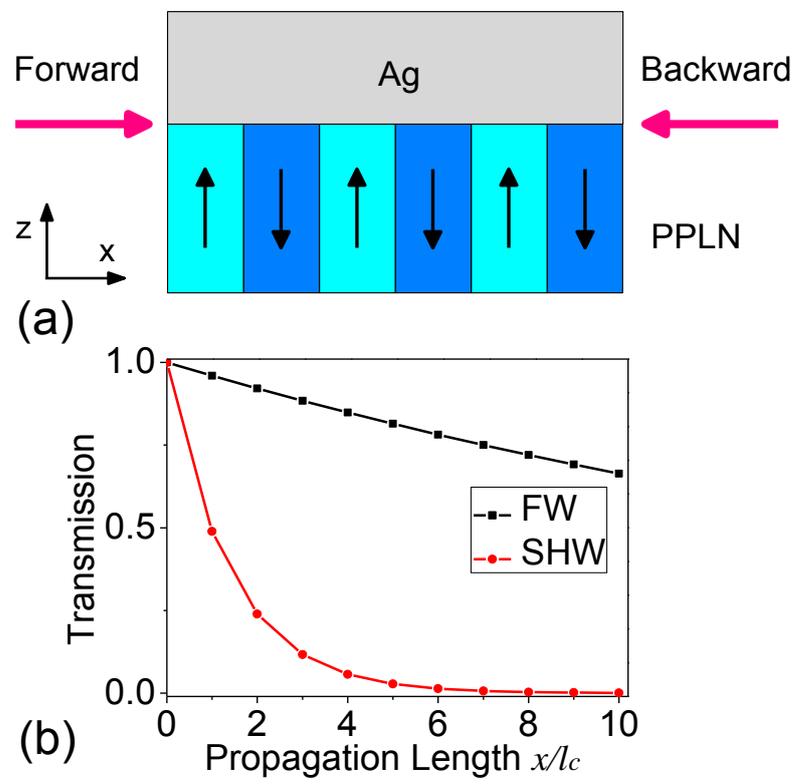

Fig. 1.

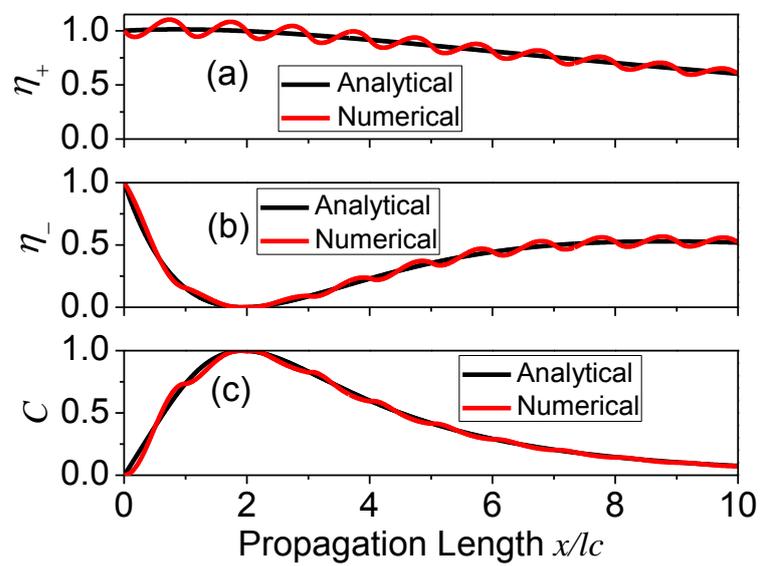

Fig. 2

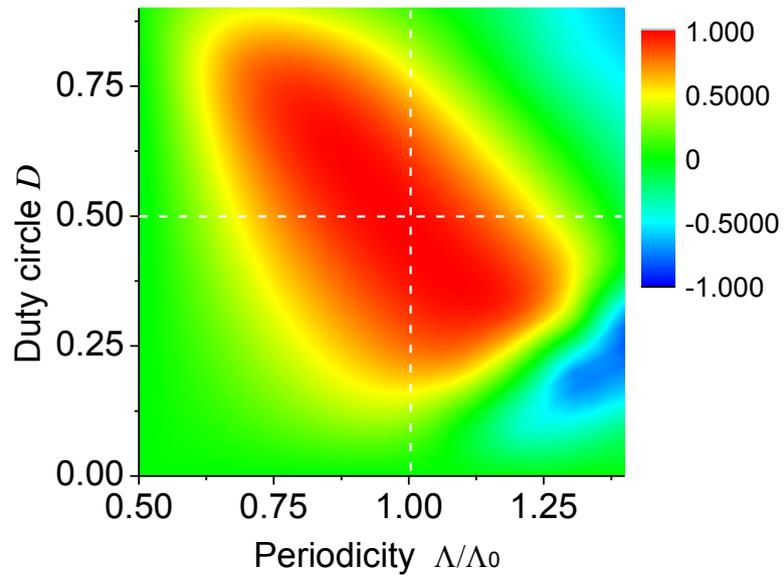

Fig.3

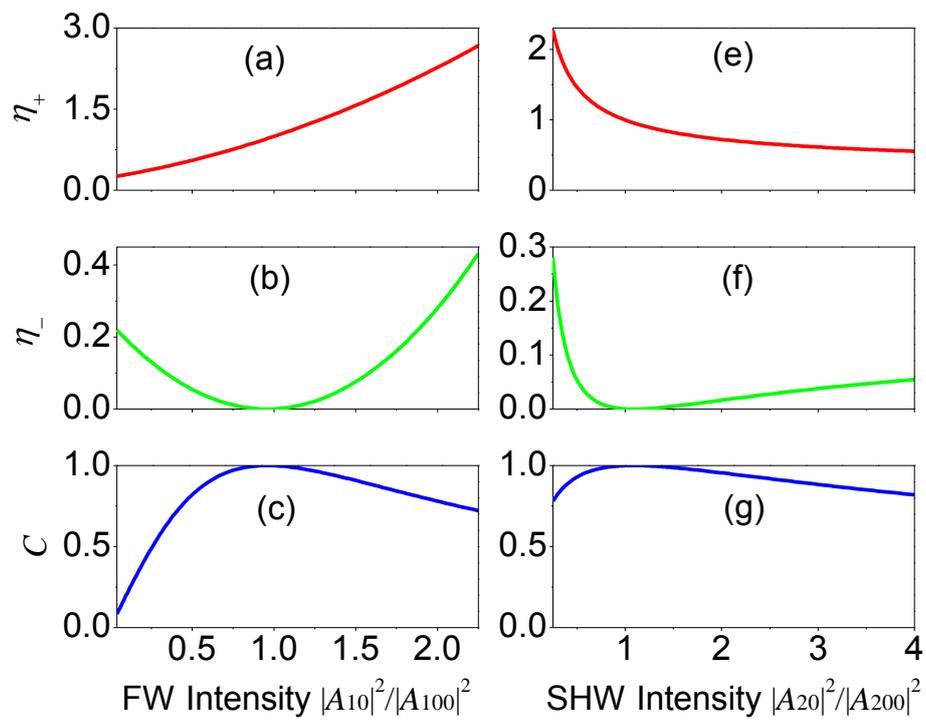

Fig. 4

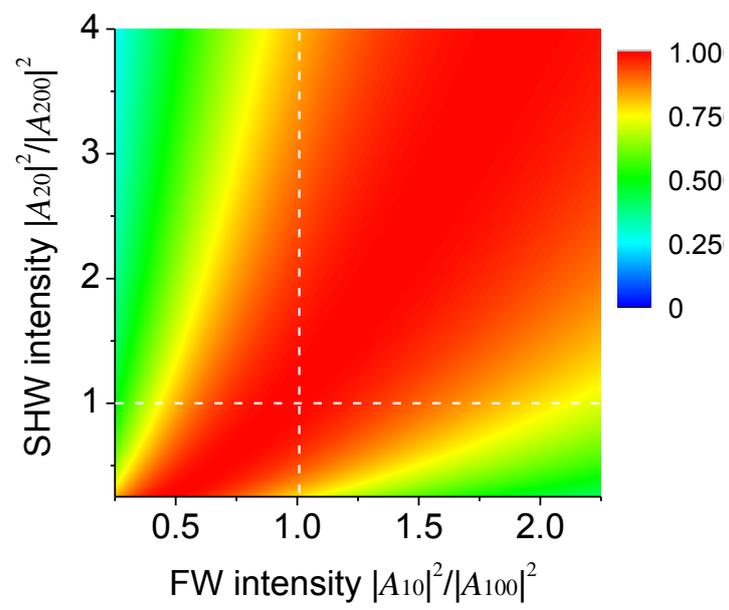

Fig. 5

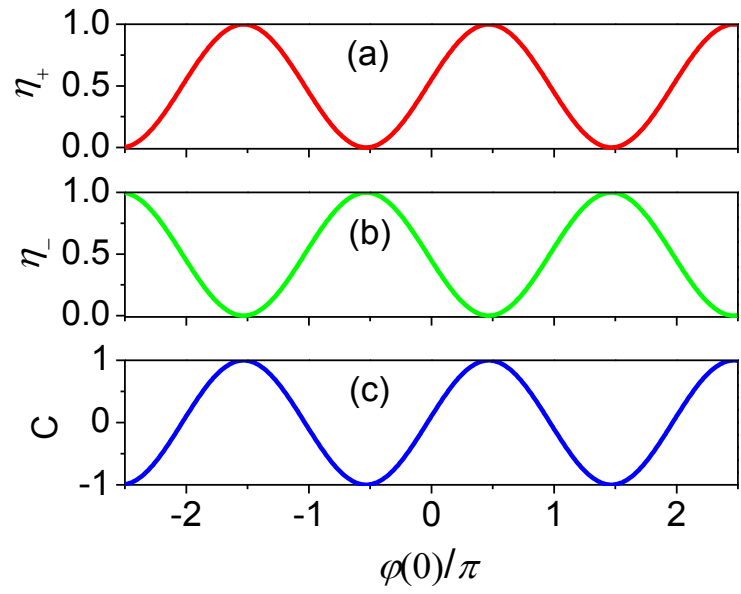

Fig. 6

| Input | | Output |
|---|---|---|
| FW | SHW | NOT SHW |
| 1 | 0 | 1 |
| 1 | 1 | 0 |

Table 1